\documentclass[reprint,superscriptaddress,amsmath,amssymb,aps,prl]{revtex4-1}

\usepackage{xcolor}
\usepackage{graphicx}
\usepackage{dcolumn}
\usepackage{bm}
\usepackage{hyperref}

\begin{document}

\preprint{APS/123-QED}

\title{Ambipolar magneto-optical response of ultra-low carrier density topological insulators}

\author{Dipanjan Chaudhuri}
\affiliation{Department of Physics and Astronomy,  The Johns Hopkins University, Baltimore, Maryland 21218, USA}

\author{Maryam Salehi}
\affiliation{Department of Physics and Astronomy, Rutgers, The State University of New Jersey, Piscataway, NJ 08854, USA}
\affiliation{Department of Materials Science and Engineering, Massachusetts Institute of Technology, Cambridge, Massachusetts 02138, USA}

\author{Sayak Dasgupta}
\affiliation{Department of Physics and Astronomy,  The Johns Hopkins University, Baltimore, Maryland 21218, USA}

\author{Mintu Mondal}
\affiliation{Department of Physics and Astronomy,  The Johns Hopkins University, Baltimore, Maryland 21218, USA}
\affiliation{School of Physical Sciences, Indian Association for the Cultivation of Science, Jadavpur, Kolkata 700032, India}

\author{Jisoo Moon}
\affiliation{Department of Physics and Astronomy, Rutgers, The State University of New Jersey, Piscataway, NJ 08854, USA}

\author{Deepti Jain}
\affiliation{Department of Physics and Astronomy, Rutgers, The State University of New Jersey, Piscataway, NJ 08854, USA}

\author{Seongshik Oh}
\affiliation{Department of Physics and Astronomy, Rutgers, The State University of New Jersey, Piscataway, NJ 08854, USA}

\author{N. P. Armitage}
\email{npa@jhu.edu}
\affiliation{Department of Physics and Astronomy,  The Johns Hopkins University, Baltimore, Maryland 21218, USA}

\date{\today}

\begin{abstract}
    We have investigated the THz range magneto-optical response of ultralow carrier density films of Sb$_2$Te$_3$ using time-domain THz polarimetry. Undoped Sb$_2$Te$_3$ has a chemical potential that lies inside the bulk valence band. Thus its topological response is masked by bulk carriers. However, with appropriate buffer layer engineering and chemical doping, Sb$_2$Te$_3$ thin films can be grown with extremely low electron or hole densities.  The ultralow carrier density samples show unusual optical properties and quantized response in the presence of magnetic fields. Consistent with the expectations for Dirac fermions, a quantized Hall response is seen even in samples where the zero field conductivity falls below detectable levels.  The discontinuity in the Faraday angle with small changes in the filling fraction across zero is manifestation of the parity anomaly in 2D Dirac systems with broken time reversal symmetry.
\end{abstract}

\maketitle

Massless Dirac fermions have a number of unusual properties. Notable among these is their unconventional Landau level spectrum. In two spatial dimensions (2D) the energy of the $n$-th Landau level, $E_n\propto \pm \sqrt{|n|B}$ [$n \in \mathbb{Z}$] where $B$ is the applied magnetic field. This is remarkably different from conventional 2D electron systems where $E_n \propto \left(n+\frac{1}{2}\right)B$ [$n \in \mathbb{Z}^+$]. Moreover, in massless 2D Dirac systems there exists a unique zeroth Landau level pinned to the charge neutral point. The magneto-transport of a single massless Dirac fermion therefore exhibits unusual features such as an extra phase factor in the Onsager quantization condition and a half integer quantum Hall effect when time reversal symmetry is broken by a magnetic field even in the limit of vanishing charge density.
\begin{equation}
    \sigma_{xy} = \left(n \pm \frac{1}{2}\right)\frac{e^2}{h}.
    \label{eq:sxy}
\end{equation}

Here $n$ is the integer filling factor and the $\pm$ is determined by sign of the charge carriers of the surface.  Notably Eq. 1 exhibits a jump of the conductance by $\frac{e^2}{h}$ when changing the carrier density from infinitesimally positive to infinitesimally negative.  In all real materials, massless Dirac fermions come in pairs giving a Hall effect twice that of Eq. 1.  Dirac-like dispersion of electrons is realized at low energies in a number of materials such as monolayer graphene, surface states of 3D topological insulators (TIs), etc. Although (accidental) band touching in solids are usually fragile, discrete symmetries such as sublattice symmetry in graphene and time reversal symmetry in topological insulators (TIs), protect the gapless dispersion in these materials. Hence these materials offer an experimental platform where the consequences of a relativistic dispersion can be explored. While graphene hosts two spin degenerate Dirac cones at $K$ and $K'$ points in the Brillouin zone \cite{Graphene2009RMP}, binary V-VI TI compounds such as Bi$_2$Se$_3$, Bi$_2$Te$_3$, Sb$_2$Te$_3$, etc. which are ideally gapped insulators in the bulk, have a single Dirac cone at the zone center on their surfaces \cite{zhang2009topological, hsieh2009observation}. These TIs are thus ideal candidates to study the unusual properties of the Landau spectrum of the Dirac fermion.  An alternative formulation of the effects of time reversal symmetry breaking on the 2D surface conduction is in terms of a 3D magnetoelectric effect \cite{qi2008topological,wu2016quantized,armitage2019matter}.

Experimentally accessing the low lying Landau levels in TIs has been challenging thus far, primarily due to material considerations. Defects in the crystal lattice tend to dope the samples such that the Fermi level is away from the Dirac node. Bi$_2$Se$_3$ for example is generally n-doped with its Fermi level above the Dirac node in the bulk conduction band whereas Sb$_2$Te$_3$ are typically p-doped with the Fermi level in the bulk valence band \cite{ginley2016topological}. This makes it challenging to study surface state dynamics through bulk probes such as transport and optical experiments and also limit practical applications. Bulk insulating high quality (mobility, $\mu\sim 1000$ cm$^2$/Vs) Bi$_2$Se$_3$ films have been grown using molecular beam epitaxy (MBE) via buffer layer engineering with sheet carrier density $\sim 10^{12}$ cm$^{-2}$ \cite{Koirala2019PRB}. More recently, we have successfully managed to grow high quality bulk insulating Sb$_2$Te$_3$ thin films using MBE via chemical doping and buffer layer engineering with exceptionally low sheet ambipolar carrier densities $\sim 10^{11}$ cm$^{-2}$, which is a few times lower than the best results obtained in the Bi$_2$Se$_3$ system \cite{salehi2019quantum}. This offers us an unique experimental platform to study the quantum Hall effect and in particular, properties of the zeroth Landau level in topological insulators which has been difficult to access without extremely high magnetic fields ($\sim$30 T).

Here we study the magneto-optical response of these films at THz frequencies to observe the topological magnetoelectric effect \cite{qi2008topological, essin2009magnetoelectric} in these ultra-low carrier density TI films. The optical response of the doped Sb$_2$Te$_3$ films with the Fermi level tuned close to the Dirac node is qualitatively different from their undoped counterpart where the Fermi level lies in the bulk valence band. Although the optical conductivity of the ultra-low carrier density films appear to be insulating, they still exhibit the quantized magnetoelectric response that is expected from the TI surface states. Moreover, because of the extremely low carrier densities, we observe the quantization in the zeroth Landau level at fields as low as 3 T. We can also distinguish the electron-hole splitting of this otherwise parity symmetric level as we approach the Dirac node from either the electron and hole sides.

\begin{figure}[t]
    \includegraphics[width = 1\columnwidth]{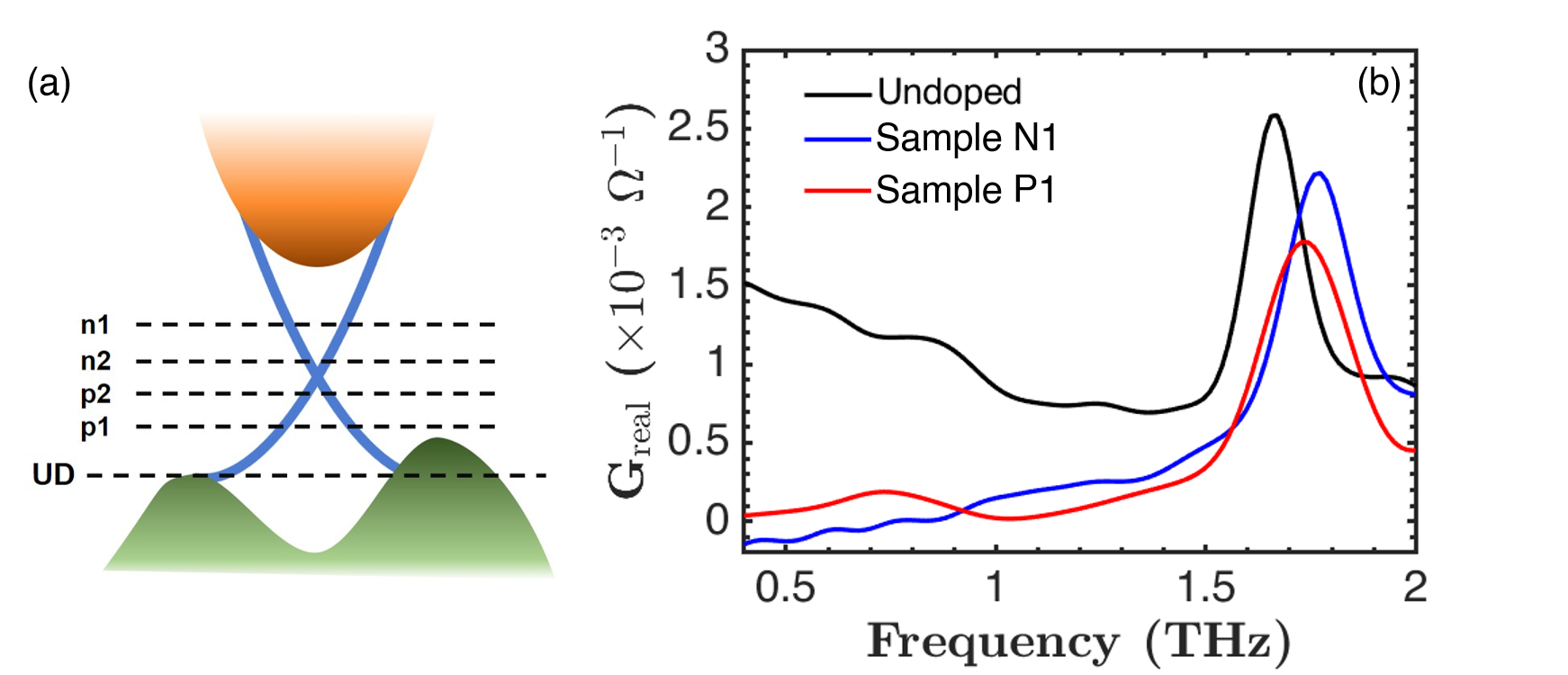}
    \caption{(a) Schematic diagram of the band structure of Sb$_2$Te$_3$ showing the Fermi level of the samples (not to scale). (b) The real part of the optical conductivity of the undoped and Ti-doped samples.}
    \label{fig:1}
\end{figure}

We measured MBE grown thin films of Sb$_2$Te$_3$ on 1 mm thick sapphire (Al$_2$O$_3$) (0001) substrates. Sb$_2$Te$_3$ films grown directly on sapphire are ordinarily p-type with a carrier density greater than $\sim$ 10$^{13}$ cm$^{-2}$ with a Fermi energy lying in the bulk valence band. However using optimally designed insulating buffer layers of 20 QL In$_2$Se$_3$ and 15 QL (In$_{0.35}$Sb$_{0.65}$)$_2$Te$_3$, we have reduced the epitaxial mismatch between Sb$_2$Te$_3$ and Al$_2$O$_3$ which subsequently reduces the carrier density. We can further tune the Fermi level across the surface Dirac node by adding $<$ 2$\%$ Ti as n-type compensation dopant and minutely adjusting its concentration. Samples were capped with a few nm thick (In$_{0.35}$Sb$_{0.65}$)$_2$Te$_3$ layer for protection and to ensure the symmetry of the environments of the top and bottom layers. More details on the sample growth can be found in Ref. \cite{salehi2019quantum}. The buffer and capping layers used are thin and insulating, and thus do not affect our optical measurements. Fig. \ref{fig:1}(a) shows a schematic diagram of the band structure of Sb$_2$Te$_3$ along with the Fermi level of the different samples studied in this letter (not to scale). Physical parameters of the Ti doped Sb$_2$Te$_3$ films measured in this letter are given in the table below. Carrier density ($n_{2D}$) and mobility ($\mu$) for the Ti doped samples were obtained from dc transport and Hall measurements at 7K. The parameters for the undoped sample are extracted from the THz spectra.

\begin{center}
    \begin{tabular}{c c c c} 
        \hline
        \hline
        Sample & Thickness & $n_{2D}$ & $\mu$ \\
        (doping) & (QL) &  (cm$^{-2}$) & (cm$^2$/Vs) \\
        \hline
        UD (undoped) & 10 & $7.2\times10^{12}$ & 1506 \\ 
        P1 (1.0\% Ti) & 6 & $2.8\times10^{11}$ & 1860 \\ 
        P2 (1.1\% Ti) & 6 & $1.7\times10^{11}$ & 1818 \\ 
        N2 (1.3\% Ti) & 10 & $-2.6\times10^{11}$ & 1527 \\
        N1 (1.5\% Ti) & 10 & $-4.5\times10^{11}$ & 1035 \\
        \hline
        \hline
    \end{tabular}
    \label{tab:1}
\end{center}

Using time-domain THz spectroscopy, we measure the complex optical conductivity of the samples between 0.3-2.1 THz. The sample was oriented such that the $c$-axis is along the direction of propagation of light. In addition, we perform high precision measurements of optical polarization rotations in magnetic field at these frequencies using a time-domain THz polarimetry setup \cite{morris2012polarization}. The real part of the Faraday rotation is the inductive response while the imaginary part encodes the dissipative response. 

Even at zero magnetic field, the optical response of the undoped and Ti-doped samples with ultralow carrier density ($\sim 10^{11}$  cm$^{-2}$) are remarkably different. Fig. \ref{fig:1}(b) compares the real part of the optical conductivity of the undoped sample with two of the Ti-doped samples. The response of the undoped Sb$_2$Te$_3$ is similar to that of the other isostructural TIs such as Bi$_2$Se$_3$, where a free electronic (Drude) contribution along with a phonon has been observed within this spectral bandwidth \cite{PhysRevLett.108.087403}. The strong absorption at 1.68 THz is due to the IR active optical phonon corresponding to the $E_u$ mode \cite{richter1982anisotropy}. In these samples with ultralow carrier density we consistently observe no signatures of Drude transport and see only conductance that is an increasing function of frequency. This is a usual signature of localization, which would be at odds with the usual understanding of topological surface state transport, which evade localization by not allowing 180$^{\circ}$ backscattering if time-reversal symmetry is preserved.  It may that the increasing function of frequency is due to ``\textit{incipient}'' localization of the surface states due to low carrier density and high disorder, which shows many of the signatures of localization while not having a dc conductance that goes completely to zero as T goes zero. We could also be observing residual impurity conduction from the bulk of the material.  Irrespective of its origin the doped low carrier density samples show no signatures for conventional metallic transport at low T.

\begin{figure*}[]
    \includegraphics[width = 2\columnwidth]{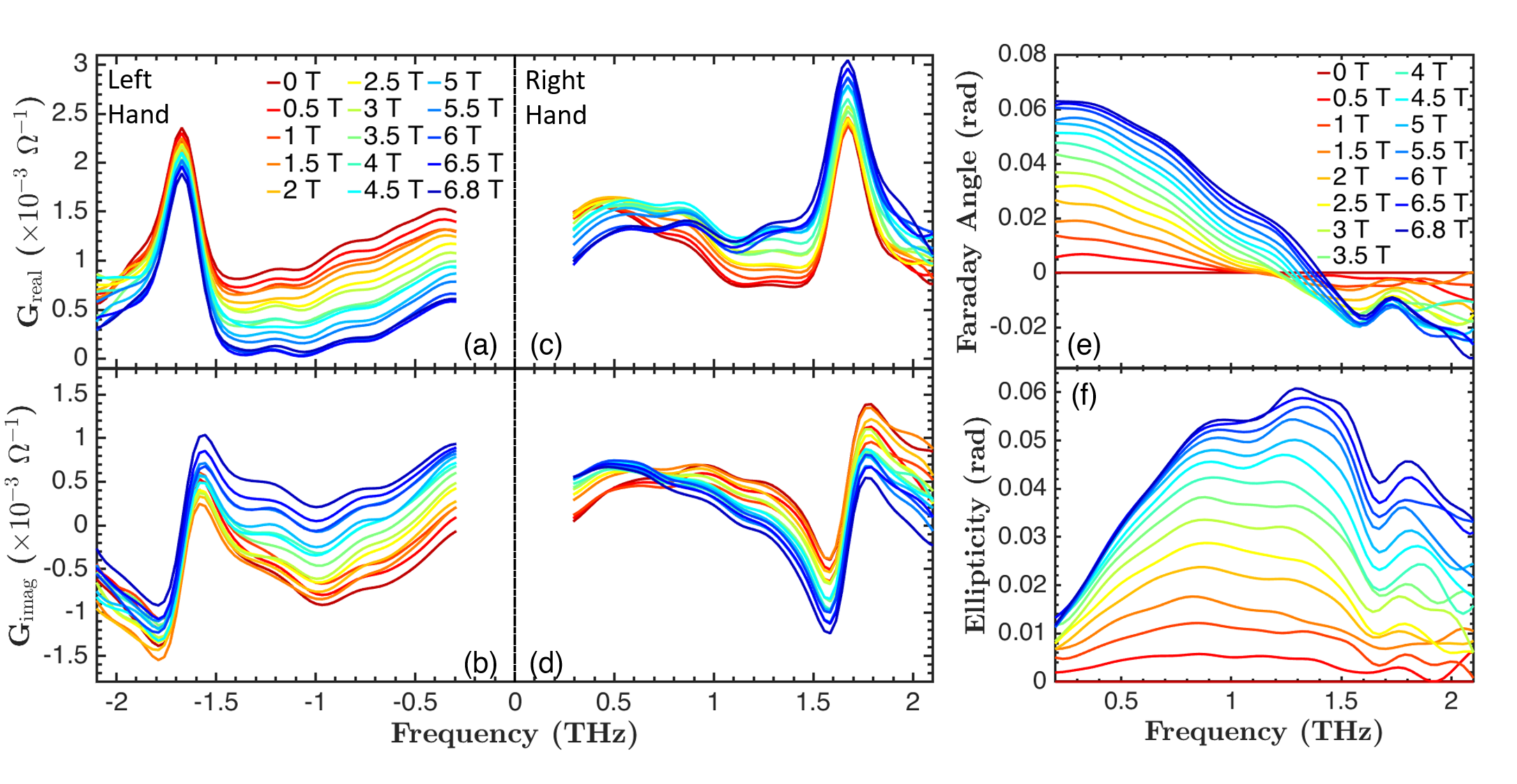}
    \caption{Real (a) and imaginary (b) part of the optical conductance of the undoped sample for different fields at 5K in the left circular basis. Panels (c) and (d) show the same in right circular basis. (e-f) Real and imaginary parts of Faraday rotation.}
    \label{fig:2}
\end{figure*}

The magneto-optical response of the undoped Sb$_2$Te$_3$ is also typical of TIs in the semiclassical regime. Fig. \ref{fig:2} (a-d) shows the measured real and imaginary parts of the optical conductivity of the undoped Sb$_2$Te$_3$ films in left and right circular basis which are the eigenpolarization in Faraday geometry (${\bf B} \parallel {\bf k}_{\text{THz}}$) \cite{armitage2014constraints}. The spectra corresponding to the left (right) circular polarization are plotted as a function of negative (positive) frequency.  With increasing magnetic field the Drude peak shifts towards positive frequencies due to the appearance of cyclotron orbits which selectively absorb one circular polarization more than the other depending on the sign of the charge carrier. As a result of this asymmetry, incident linearly polarized light becomes elliptically polarized upon passing through this sample. This is a typical Faraday effect. Fig. \ref{fig:2} (e-f) shows the measured complex Faraday rotation where the real and imaginary parts correspond to the rotation and ellipticity of the polarization. The sign of the rotation together with the direction of the shift of the Drude peak indicate p-type carriers, consistent with what is expected in undoped Sb$_2$Te$_3$. As the Fermi level is embedded within the bulk valence band, this response is due to the contribution from both bulk and surface state carriers. Hence we do not observe quantized Faraday rotation which is expected in case of pure surface state transport.  

Despite the lack of signatures of free carriers in the zero field optical conductivity in the doped samples, the Faraday rotation is remarkably consistent with integer quantized magnetoelectric response of the surface states \cite{qi2008topological, essin2009magnetoelectric}. Sb$_2$Te$_3$ hosts a single Dirac fermion on its surface and as such should exhibit a half-integer quantized Hall effect [Eq. \ref{eq:sxy}], a consequence of the $\pi$ Berry phase carried by the Dirac fermion. However in a finite 3D sample, there is an added phase factor from the curvature of the 2D space which forms the genus $g = 0$ boundary of the 3D sample. Therefore, it is not possible to observe this half-integer quantization directly \cite{Lee2009PRL} as the same surface state is probed twice in any experiment \cite{Nagaosa2015PRL}. The quantum Hall effect in a 3D TI slab in uniform magnetic flux threading though the sample is thus given by
\begin{equation}
    \sigma_{xy} = \left(n_t \pm \frac{1}{2}+n_b \pm \frac{1}{2} \right)\frac{e^2}{h}.
    \label{eq:sxy_TI}
\end{equation}
$n_{t/b}$ represents the filling factor of the Landau levels in the top/bottom surfaces which will be measured together as discussed above.  The Faraday angle also reflects this quantization as it is independent of frequency and field for low frequencies. The quantized value is consistent with the observations of topological magnetoelectric effect in Bi$_2$Se$_3$ \cite{wu2016quantized, mondal2018electric}:
\begin{equation}
    \textbf{Re}(\theta_F) \simeq \frac{2\alpha}{n_s+1}\left(n_t \pm \frac{1}{2}+n_b \pm \frac{1}{2}\right)
\end{equation}
where $\alpha \sim 1/137.04$, is the fine structure constant and the refractive index of the substrate, $n_s = 3.1$.  The factors of $\frac{1}{2}$ are emblematic of the half-integer quantization of the individual surfaces and the sense of the Faraday rotation is determined by the sign of the charge carriers. 

\begin{figure*}[]
    \includegraphics[width = 2\columnwidth]{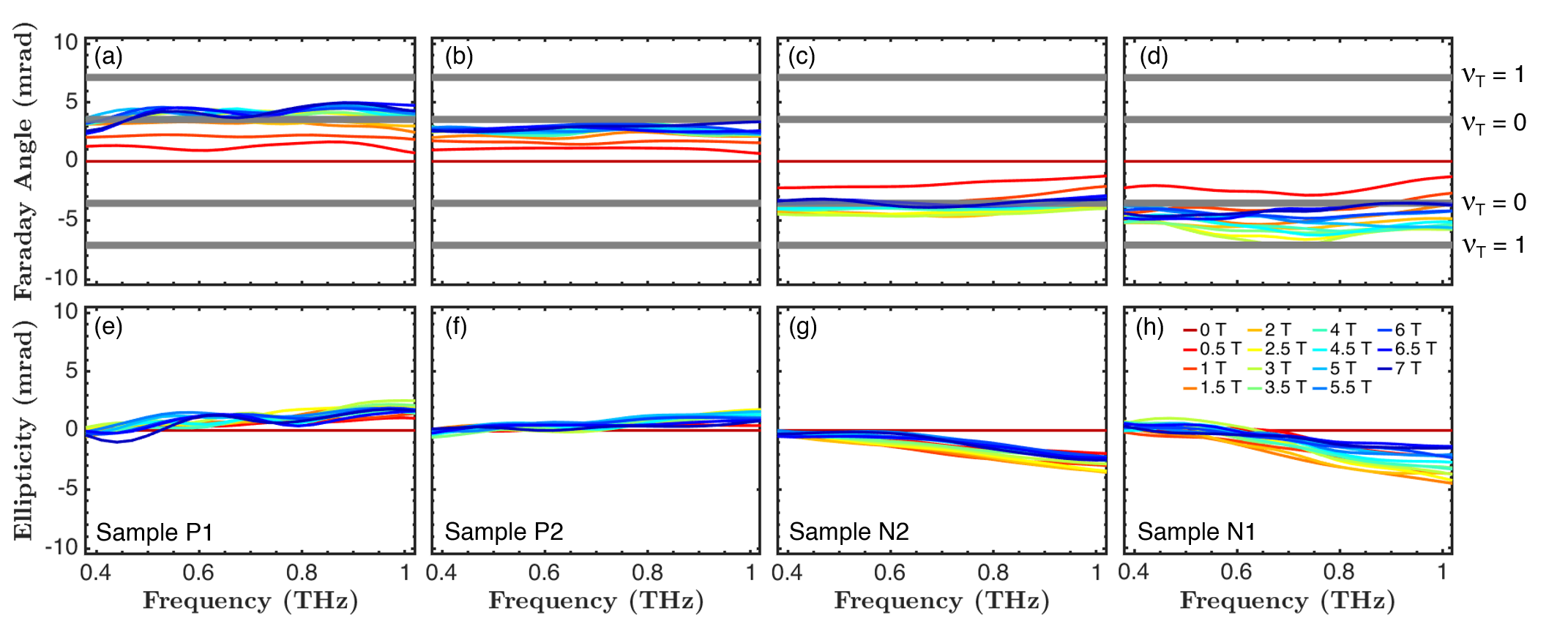}
    \caption{Real [imaginary] parts of the Faraday rotation for different magnetic fields for four different samples with hole (a, b) [(e, f)] and electronic (c, d) [(g, h)] charge carriers. Grey solid lines represent the expected quantized Faraday angle for filling factors $n = 0, 1$ for both hole and electronic carriers.}
    \label{fig:3}
\end{figure*}

\begin{figure*}
    \centering
    \includegraphics[width = 1.5\columnwidth]{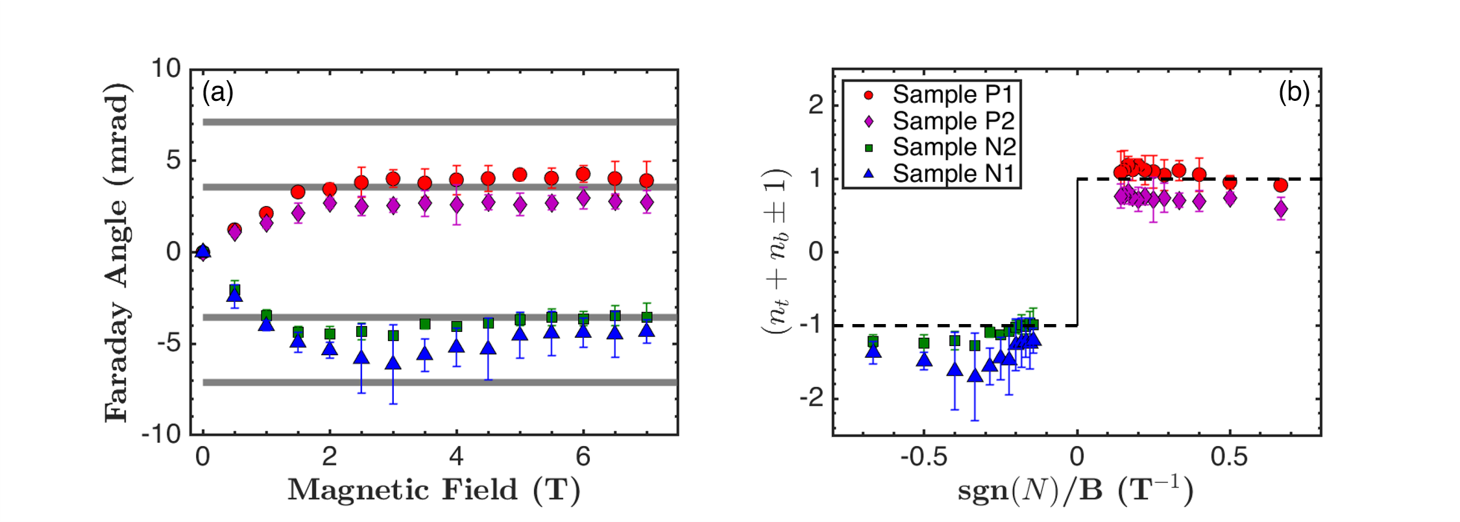}
    \caption{(a) Real part of the average Faraday rotation as a function of magnetic field for the Ti-doped samples. Grey solid lines represent the expected quantized Faraday angles ($n = 0, 1$). (b) Average Faraday angle as a function of inverse magnetic field. Black dashed line is a guide to eye for the expected evolution. Black solid line marks the discontinuous jump unique to Dirac electrons.}
    \label{fig:4}
\end{figure*}

Fig. \ref{fig:3} shows the real (a - d) and imaginary (e - h) parts of the Faraday rotation as a function of frequency for different fields for four different doped samples of both polarities. The grey lines indicate the predicted values of the quantized Faraday angle for a combined filling factor of the top and bottom surface states $n_{T} = n_t + n_b$. In all cases the Faraday rotation starts at small values before increases with increasing field to approach the grey quantized $n_{T} = 0$ lines at larger fields.  This indicates that even at moderately low fields, the samples enter the quantized regime and owing to the low carrier density, the carriers are confined to the zeroth Landau level. Note from the ellipticity data that this quantized response is found despite the finite dissipative response (particularly in the n-type samples). Also note, that because of the low carrier density, the optical conductivity does not change significantly with field and thus polarimetry is a more sensitive probe of the charge dynamics in these samples.

Fig. \ref{fig:4}(a) shows the frequency averaged Faraday angle (0.4--0.8 THz) as a function of magnetic field.  The Faraday angle in the p-type samples first increase with increasing field before saturating at a quantized value near 3 T. The n-type samples however behave somewhat differently where the Faraday angle at first goes beyond the quantized value with increasing field. Upon further increase of the magnetic field, the Faraday angle reverts back to the quantized plateau around 5T and then eventually saturates.  The quantization of Faraday angle in these samples are not as precise as previously observed in Bi$_2$Se$_3$ \cite{wu2016quantized}. This may be due to a number of reasons. Firstly, there is the experimental challenge of measuring extremely small rotations (3.8 mrad $\sim$ 0.22$^{\circ}$) over a broad frequency range accurately. Secondly, light-matter interaction itself is greatly diminished as the number of carriers responding to the oscillating electromagnetic radiation is reduced. Moreover, the charge carriers are subject to a disorder potential created by random charged impurities leading to the formation of charge puddles. This has been studied extensively in other massless Dirac systems such as graphene where charge puddles grow as the Fermi level approaches the charge neutral point \cite{samaddar2016charge}.   The nature of disorder could be very different in these low carrier density systems leading to dissipation from isolated regions.

In Fig. \ref{fig:4}(b) we plot the average Faraday angle in the quantized regime (B$\geq$5T) for different films with varying carrier density as a function of inverse field (proportional to filling factor).  Despite the low carrier densities, the sample is sensitive to the sign of the charge carrier and promptly switches sign as one sweeps across the Dirac node and it is evident from the data that $\lim\limits_{n_{2D}\rightarrow0-}\theta_F \neq \lim\limits_{n_{2D}\rightarrow0+}\theta_F$. Finally, we can point out that the discontinuous jump with small doping change in the Faraday angle in Fig. 4(b) is a manifestation of the parity anomaly in 2D Dirac systems with broken time reversal symmetry \cite{Oleg2000Parity}. Our experiments are consistent with the expected discontinuity. 

\begin{acknowledgments}

This work at JHU and Rutgers was supported by MURI W911NF2020166.   We would like to thank O. Tchernyshyov for helpful conversations.

\end{acknowledgments}


\bibliography{references}

\end{document}